\documentclass{article}
\usepackage[english]{babel}
\usepackage[margin=0.8in]{geometry}
\usepackage{multicol}
\usepackage{multirow}
\usepackage{background}
\usepackage[T1]{fontenc}
\usepackage[utf8x]{inputenc}
\usepackage[math]{blindtext}
\usepackage{amsmath}
\usepackage{breqn}
\usepackage{cite}
\usepackage{mathptmx}
\usepackage{anyfontsize}
\usepackage{t1enc}
\usepackage{mathalfa}
\usepackage{amsfonts}
\usepackage{amssymb}
\usepackage{array}
\usepackage{flushend}
\usepackage{placeins}
\usepackage{graphicx}
\usepackage{bm}
\usepackage[colorlinks]{hyperref}
\usepackage{url}
\usepackage{color}
\definecolor{darkred}{rgb}{0.5,0,0}
\definecolor{darkgreen}{rgb}{0,0.5,0}
\definecolor{darkblue}{rgb}{0,0,0.5}
\hypersetup{
  colorlinks,
  citecolor=darkblue,
  linkcolor=darkred,
  urlcolor=darkblue}
\usepackage{float}
\usepackage{lipsum}

\SetBgScale{4}
\SetBgColor{gray}
\SetBgAngle{90}
\SetBgContents{}
\SetBgPosition{-1.5,-10}

\begin{document}

{\centering

{\bfseries\Large Complete One-Loop Corrections to $e^+ e^-\rightarrow{\tilde{\chi}}_1^0 {\tilde{\chi}}_1^0 h^0$ for Different Scenarios\bigskip}

S. M. Seif\textsuperscript{1} , T.A. Azim\textsuperscript{1, 2}\\
  {\itshape
\textsuperscript{1}Faculty of Science, Physics Department, Cairo University, Giza, Egypt. \\
\textsuperscript{2}Faculty of Science, Physics Department, Albaha University , Albaha, KSA. \\

  }
}

\begin{abstract}
In this work, the radiative corrections to the production of a light Higgs boson ($h^0$)  
with a pair of lightest neutralinos (${\tilde{\chi}}_1^0$) in $e^+ e^-$ collisions within MSSM are presented, including the on-shell renormalization scheme in the loop calculations. We have studied the QED corrections as well as the weak corrections, where the contribution from both corrections is significant and needs to be taken into account in the future linear colliders experiments. The result includes the numerical calculations for two different SUSY scenarios---Higgsino and Gaugino scenarios--- for $e^+ e^-\rightarrow{\tilde{\chi}}_1^0 {\tilde{\chi}}_1^0 h^0$.
\bigskip

\end{abstract}

\begin{multicols}{2}
\section{Introduction}
\label{intro}
Linear electron-positron colliders are considered to be the best environment for precise studies of supersymmetric models, especially for the Supersymmetric Standard Model (MSSM). There are five Higgs mass states in MSSM: two CP-even, \textit{h}$^{0}$ and \textit{H}$^{0}$, a CP-odd, \textit{A}$^{0}$ and a pair of charged bosons, \textit{H}$^{\pm}$. As clarified in \cite{r1,r2}, all Higgs bosons in the MSSM, except the lightest CP-even one, are too heavy to play an important role in the current and the near future experiments. Therefore, the present study concentrates on the lightest Higgs boson \textit{h}$^{0}$ only. On July 2012 \cite{r3,r4,r5}, ATLAS and CMS teams at LHC have announced independently the discovery of a boson with the similar properties of that of Higgs boson and confirmed likely, on March 2013, to be a Higgs boson of mass $\thicksim 125$ GeV. 
This major breakthrough has a great impact on the searching for other supersymmetric particles and the mechanism of the supersymmetry (SUSY) breaking. The lightest MSSM CP-even Higgs particle mass is bounded from above and, depending on the SUSY parameters $M_A$ and $\tan\beta$, is in the range of $m_h^{max} \thickapprox 90-130$ GeV. The lower value comes from experimental constraints at LEP \cite{r6,r7}, while the upper bound assumes a SUSY breaking scale $M_s \lesssim \mathcal{O}(1 TeV)$.   

The mass of neutralinos is among the precision observables with lots of information on the SUSY-breaking structure, the relations between the particle masses and the SUSY parameters are important theoretical quantities for precision calculations. In MSSM \cite{r8}, one has four neutralinos $\tilde{\chi}_1^0$-$\tilde{\chi}_4^0$, which are the fermion mass eigenstates of the supersymmetric partners of the photon, the $Z^0$ boson, and the neutral Higgs bosons $\textit{H}_{1,2}^0$. Their mass matrix depends on the parameters $M_1$, $M_2$, $\mu$, and $\tan\beta$. If SUSY is realized in nature, neutralinos should be found in the present high energy experiments at Tevatron, LHC \cite{r9} and future $e^+ e^-$ colliders. Especially at a linear $e^+ e^-$ collider, it will be possible to perform measurements with high precision \cite{r10,r11}. 

To get high matching between the experimental predictions and theoretical calculations, it is unavoidable to include higher-order terms. In this paper, we use on-shell renormalization scheme in the loop calculations of the Higgs, neutralino sectors, and all SUSY particles of the CP-conserving MSSM. The calculation was performed using the FeynArts-3.6, FormCalc-7.1 and LoopTools-2.7 packages. At the one-loop level calculations, we have implemented all the renormalization constants, required to determine the various counterterms for the Higgs, neutralino and other sectors in the MSSM model file of FeynArts \cite{r12}. FormCalc was used to algebraically simplify the resulting 
amplitudes, which were converted to a FORTRAN program for integral evaluation using LoopTools. The corrections due to the real photon emission are calculated to cancel the IR singularities present in virtual corrections at one-loop level.

The paper is organized as follows: The analytical calculations of the electroweak radiative corrections to the $e^+ e^-\rightarrow{\tilde{\chi}}_1^0 {\tilde{\chi}}_1^0 h^0$ process is given in section ~\ref{sec:1}, involving the soft photonic corrections. The numerical results are presented in ~\ref{sec:2}. Finally, the conclusions are given in ~\ref{sec:3}.

\section{Radiative Corrections}
\label{sec:1}

The associated production of MSSM neutral Higgs bosons with neutralinos is liklely substantial due to the large dependence of their coupling on the soft–SUSY breaking parameters, which are important theoretical quantities for precision calculations, and subsequently carry information on the SUSY theory \cite{r13}. 

\begin{figure}[H]
\includegraphics{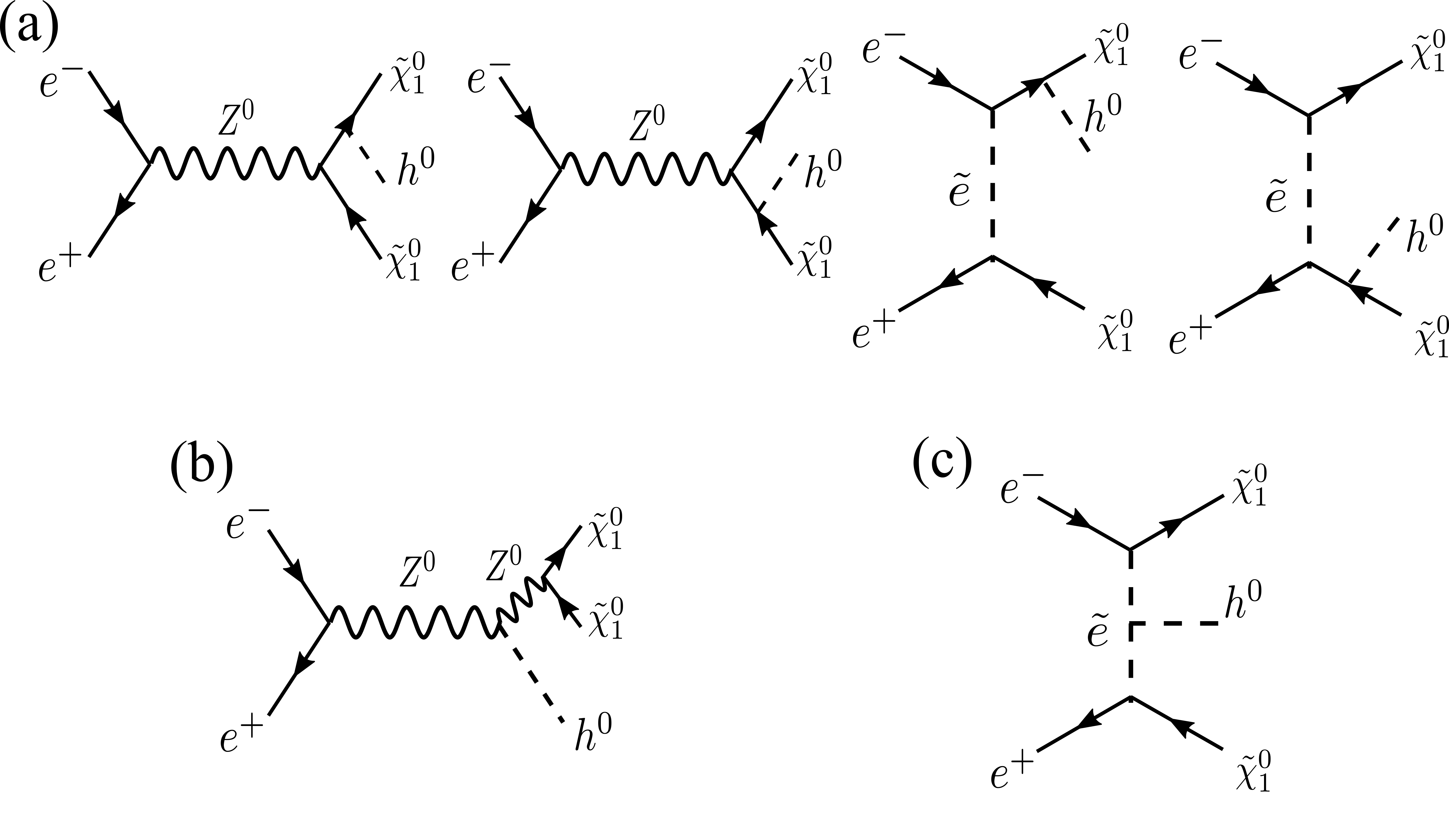}
\caption{\label{fig:1}The lowest order (LO) Feynman diagrams for the $e^+ e^-\rightarrow{\tilde{\chi}}_1^0{\tilde{\chi}}_1^0 h^0$ process.}
\end{figure}

The LO predictions for the cross sections suffer from large uncertainties because LO calculations has less precise results, so the higher order should be included. This process which is written as: 

\begin{eqnarray}
e^+(p_1)+e^-(p_2)\rightarrow{\tilde{\chi}}_1^0(p_3)+{\tilde{\chi}}_1^0(p_4)+h^0(p_5)\nonumber
\end{eqnarray}
at the tree level are described by the Feynman diagrams of Fig.~\ref{fig:1}. The momenta of the particles are given in brackets. The momenta obey the on-shell conditions $p_1^2=p_2^2=0$, $p_3^2=p_4^2=m_{{\tilde{\chi}}_1^0}^2$, $p_5^2=m_h^2$.The center-of-mass energy squared $s=(p_1+p_2)^2$. The tree-level total cross section for this process can be written as:

\begin{eqnarray}
\sigma^0 (e^+ e^-\rightarrow{\tilde{\chi}}_1^0{\tilde{\chi}}_1^0 h^0)= \frac{(2\pi)^{4}}{4 |\overline{p}_1| \sqrt{s}} \int\sum\limits_{spins}{|\mathcal{M}_0|^2}d\Phi_3.
\end{eqnarray}

where $d\Phi_3$ is the three-body phase space element: 

\begin{eqnarray}
d\Phi_3 = \delta^4(p_1+p_2-\sum\limits_{i=3}^5p_i)\prod\limits_{j=3}^5 \frac{d^{3}P_j}{(2\pi)^{3}2E_j},\nonumber
\end{eqnarray}

Feynman diagrams in Fig.~\ref{fig:1} represent the 6 most contributing topologies involved in this process:
\begin{itemize}
  \item 3 with the s-channel $Z^0$ exchange,
  \item 3 with the t-channel left– and right–handed selectron ${\tilde e}_{L,R}$ exchange.
\end{itemize}
The diagrams where the $h^0$ boson is emitted from the electron and positron lines give negligible contributions. 

For high precise results, radiative corrections should be included in the calculations of the total cross setion which involve virtual one-loop correction and real photon emission such that: 

\begin{align}
\label{eq:2}
\sigma^{total} &= \sigma^0 + \delta\sigma,
\end{align}
where

\begin{align}
\label{eq:3}
\delta\sigma  &= \sigma^{virt} + \sigma^{real}\\
&= \int\sum\limits_{spins}{|\mathcal{M}_{virt}(e^+ e^-\rightarrow{\tilde{\chi}}_1^0{\tilde{\chi}}_1^0 h^0)|^2}d\Phi_3\nonumber\\
&+\int\sum\limits_{spins}{|\mathcal{M}_{real}(e^+ e^-\rightarrow{\tilde{\chi}}_1^0{\tilde{\chi}}_1^0 h^0\gamma)|^2}d\Phi_4\nonumber.
\end{align}

\begin{figure}[H]
\includegraphics[width=.49\textwidth]{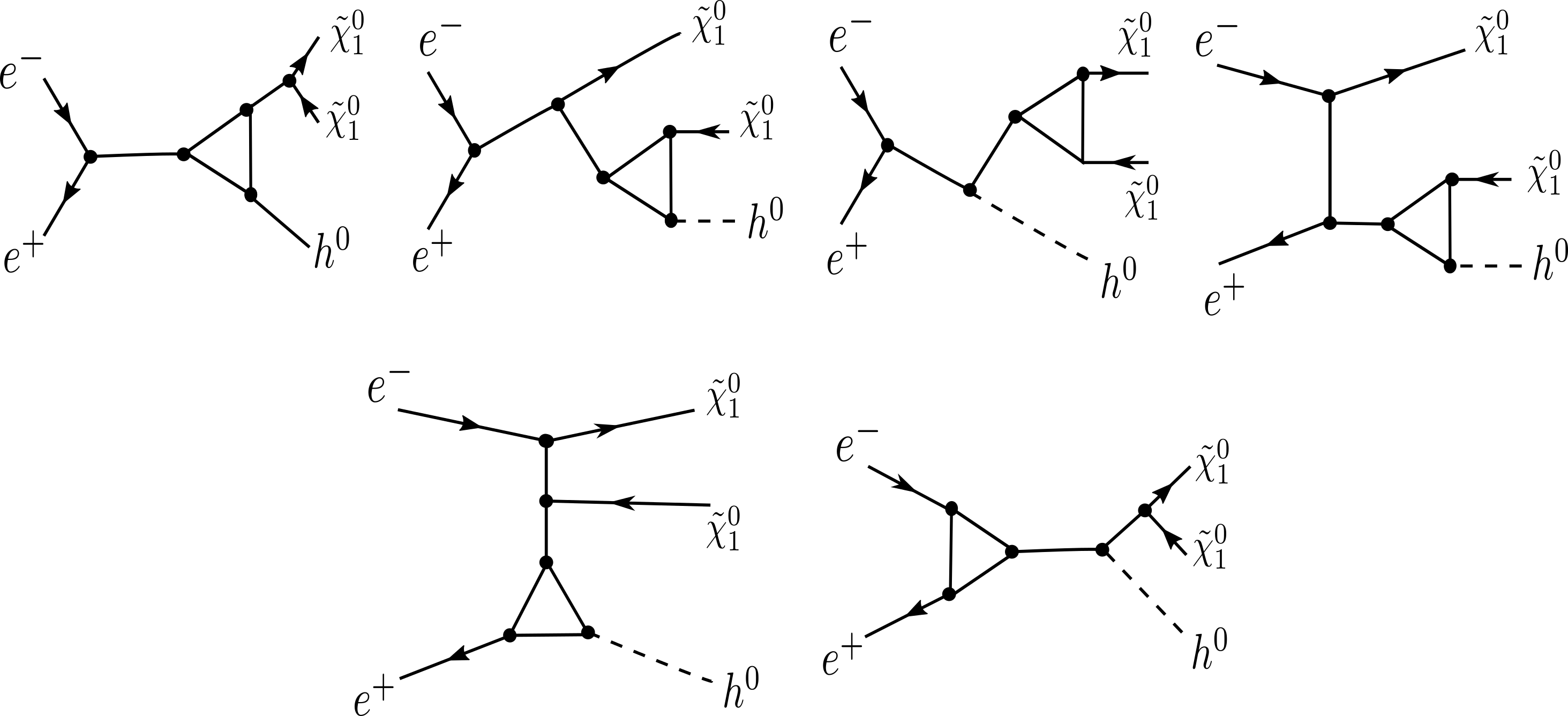}
\caption{\label{fig:2}Vertex Corrections.}
\end{figure}

\begin{figure}[H]
\includegraphics{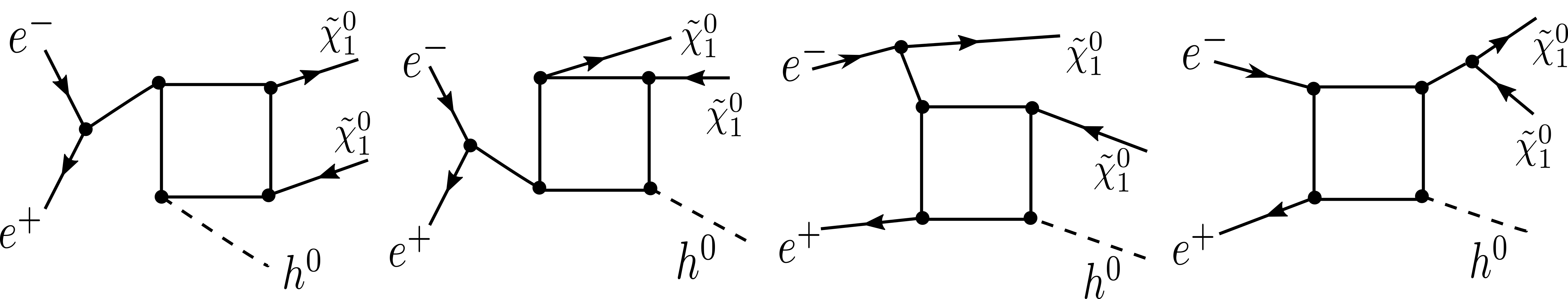}
\caption{\label{fig:3}Box Corrections.}
\end{figure}

\begin{figure}[H]
\includegraphics[width=.5\textwidth]{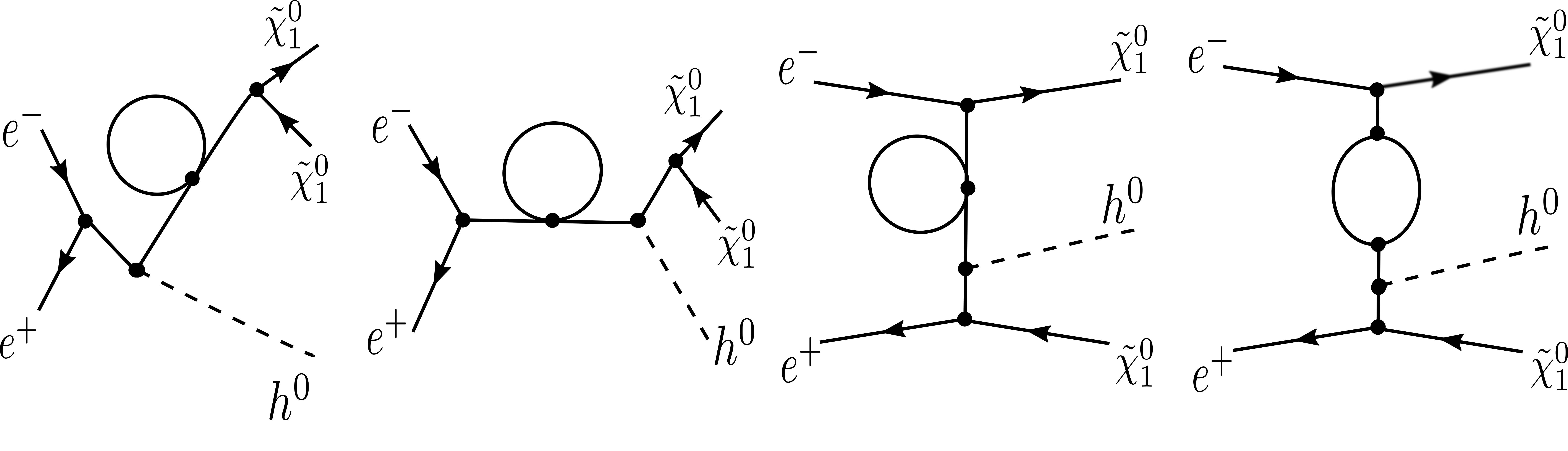}
\caption{\label{fig:4}Propagator Corrections.}
\end{figure}

One-loop Feynman diagrams can be classified as the following generic structure: The virtual vertex corrections Fig.~\ref{fig:2}, the box graph contributions to the propagators Fig.~\ref{fig:3}, and the self-energy contributions Fig.~\ref{fig:4}. The complete supersymmetric spectrum is used for the virtual particles inside loops. The evaluation of one-loop diagrams usually leads to two types of divergences:    
\begin{itemize}
  \item UV divergences, which are associated with singularities occurring at large loop momenta,
  \item IR divergences, which are generated, if one of the propagators in the loop vanishes.
\end{itemize}
To isolate the UV divergences, the regularization by dimensional reduction scheme (DR) is used to preserve SUSY. In this scheme only the momenta are treated as \textit{D}-dimensional, while the fields and the Dirac algebra are kept 4-dimensional. To get rid of the UV divergences and absorb them, they should be renormalized by introducing a suitable set of counterterms for the renormalization of the coupling constants and the renormalization of the external wave functions. In this paper on-shell renormalization scheme is used in which all particle masses are defined as pole masses, such that the cross sections are directly related to the physical masses of the external particles and the other particles entering the loops \cite{r14,r15}. The complete cross section at the one-loop level can be written as follows:

\begin{align}
\label{eq:4}
\sigma^{1-loop}&=\sigma^0+\sigma^{virt}
\end{align}

The virtual electroweak radiative correction to the cross section is given by:

\begin{align}
\label{eq:5}
\sigma^{virt}&=\sigma^0\Delta_{virt}=\frac{(2\pi)^{4}}{2 |\overline{p}_1| \sqrt{s}} \int d\Phi_3\sum\limits_{spins}\Re(\mathcal{M}^{\dagger}_{0}\mathcal{M}_{virt})
\end{align}
where $\Delta_{virt}$ is the relative virtual correction and  $\mathcal{M}_{virt}$ is the renormalized amplitude involving all the one-loop electroweak Feynman diagrams and corresponding counterterms.
The contributions of virtual photon exchange in loops leads to soft IR divergences as well as the real photon emission \cite{r14}, but their sum is IR finite. 

From previous discussion the corrected cross section can be expressed as following:

\begin{align}
\sigma^{corr}&(e^+ e^-\rightarrow{\tilde{\chi}}_1^0{\tilde{\chi}}_1^0 h^0)\nonumber\\ 
&= \sigma^{ren}(e^+ e^-\rightarrow{\tilde{\chi}}_1^0{\tilde{\chi}}_1^0 h^0) + \sigma^{ren}(e^+ e^-\rightarrow{\tilde{\chi}}_1^0{\tilde{\chi}}_1^0 h^0\gamma)\nonumber.
\end{align}

\subsection{\label{sec:level2}Renormalization of Neutralino Sector}

The tree level neutralino mass terms are given by:

\begin{eqnarray}
\mathcal{L}_n = -\frac{1}{2}[\psi^{0^\top}Y\psi^0 + \bar{\psi}^{0^\top}Y^\dagger\bar{\psi}^0] + \textit{h}.\textit{c}.\nonumber,
\end{eqnarray}
where

\begin{eqnarray}
\psi^0 =(\tilde{B}^0, \tilde{W}^3, \tilde{h}_1^0, \tilde{h}_2^0)^T.\nonumber
\end{eqnarray}
Lagrangian involves the $\mu$ parameter, the soft-breaking gaugino-mass parameters $M_1$ and $M_2$, and the Higgs vacua $v_i$, which are related to $\tan\beta=v_2/v_1$ and to the 
$W$ mass $M_W=gv/2$ with $(v_1^2+v_2^2)^{1/2}$ \cite{r15,r17}.

After the electroweak symmetry is broken, the neutralino mass matrix in the bino–wino–higgsino basis can be written as: 

\begin{frame}
\tiny
\arraycolsep=0.5pt 
\medmuskip = 1mu 
\[ 
\begin{split}
Y&=\\
&\left(\begin{array}{cccc}
M_1 & 0 & -M_Zs_W\cos\beta & M_Zs_W\sin\beta\\
0 & M_2 & M_Zc_W\cos\beta & -M_Zc_W\sin\beta\\
-M_Zs_W\cos\beta & M_Zc_W\sin\beta & 0 & -\mu\\
M_Zs_W\cos\beta & -M_Zc_W\sin\beta & -\mu & 0
\end{array}\right),\end{split}\]
\end{frame}\\
which can be diagonalized with the help of a unitary $4\times4$ matrix $N$, yielding the neutralino mass eigenstates $\tilde{\chi}_i^0(i=1,\ldots,4)$.

Renormalization constants are introduced for the neutralino mass matrix $Y$ and for the neutralino fields $\psi^0$  by the transformation:

\begin{align}
Y & \rightarrow Y+\delta Y,\nonumber \\
\psi^0 & \rightarrow (1+\frac{1}{2}\delta Z_{\tilde{\chi}^0})\psi^0.
\end{align}
where The matrix-valued renormalization constant $\delta Z_{\tilde{\chi}^0}$ is a general complex $4\times4$ matrix of one-loop order.

The physical (on-shell) masses are defined as poles of the real parts of the one-loop corrected propagators. The physical neutralino masses are then given by:

\begin{eqnarray}
\label{eq:7}
m^{os}_{\tilde{\chi}^0_i} = m_{\tilde{\chi}^0_i} +(N^*\delta Y N^-1)_{ii} - \delta m_{\tilde{\chi}^0_i},
\end{eqnarray}
where $m_{\tilde{\chi}^0_i}$ is the finite tree level mass, and $\delta m_{\tilde{\chi}^0_i}$ is the loop correction to the neutralino mass. The pole mass $m^{os}_{\tilde{\chi}^0_i}$ is considered as an input by specification of the parameters $\mu$, $M_1$, $M_2$, which are related to the input masses in the same way as in LO. In this way, the tree-level masses $m_{\tilde{\chi}^0_i}$ as well as the counterterm matrix $\delta Y$, are fixed. 

The matrix $\delta{Y}$ consists of the counterterms for the following parameters in the mass matrix Y: $M_1$, $M_2$, $\mu$, $\tan\beta$, the $Z$ boson mass $M_Z$, $W$ boson mass $M_W$, which is involved in $\theta_W$, and the electroweak mixing angle $s_W=\sin\theta_W$, $c_W=\cos\theta_W$, such that: 
\begin{frame}
\tiny
\arraycolsep=0.5pt 
\medmuskip = 1mu 
\[ 
\delta Y =
\left(\begin{array}{cccc}
\delta M_1 & 0 & \delta Y_{13} & \delta Y_{14}\\
0 & \delta M_2 & \delta Y_{23} & \delta Y_{24}\\
-\delta Y_{31} & \delta Y_{32} & 0 & -\delta \mu\\
\delta Y_{41} & -\delta Y_{42} & -\delta \mu & 0
\end{array}\right).\]
\end{frame}\\

$\delta M_W^2$, $\delta M_Z^2$ and $\delta\theta_W$ are the same as in SM. We renormalize them according to the on–shell prescription of electroweak renormalization, where $M_W$ and $M_Z$ are physical (pole) masses, and $\cos\theta_W=M_W/M_Z$. This gives \cite{r18}:

\begin{align}
\delta M_W^2 &= \tilde{\Re}\Sigma_{WW} (M_W^2 ),\nonumber\\
\delta M_Z^2 &=\tilde{\Re}\Sigma_{ZZ} (M_Z^2 ),\nonumber\\
\delta \cos\theta_W &=\frac{M_W}{M_Z}\left(\frac{\delta M_W}{M_W}-\frac{\delta M_Z}{M_Z}\right).
\end{align}
$\Sigma_{WW}$ and $\Sigma_{ZZ}$ are the transverse components of the diagonal $W$ and $Z$ two-point functions in momentum space, respectively. Those three counterterms have, besides the contributions from the SM, new contributions from the MSSM involving loops of superparticles and additional Higgs bosons. $\delta \tan \beta$ is fixed in Higgs sector as following:

\begin{eqnarray}
\delta \tan \beta = \frac{1}{2 M_Z \cos^2\beta}\Im\left(\Sigma_{AZ}\left(m_A^2\right)\right),
\end{eqnarray}
this implies that the two-point function connecting the CP-odd Higgs boson $A$ to $Z$ boson vanishes when $A$ is on-shell, where $M_A$ is the mass of the $A^0$ boson..
$\delta M_1$, $\delta M_2$ and $\delta \mu$ are fixed in neutralinos sector. By using the three neutralino masses as inputs, the counterterms $\delta M_1$, $\delta M_2$ and $\delta \mu$ are all determined from Eq.(\ref{eq:6}).

\subsection{\label{sec:level2}Renormalization of Higgs Sector}

As known the MSSM requires two Higgs doublets $H_1$ and $H_2$ with opposite hypercharge $Y_1=-Y_2=-1$. The quadratic part of the Higgs potential in the MSSM is given by:

\begin{eqnarray}
\label{eq:10}
V = m_1^2 H_1 \bar{H}_1 + m_2^2 H_2 \bar{H}_2 + m_{12}^2 (\epsilon_{ab} H_1^a H_2^b + h.c.)\nonumber\\
+ \frac{1}{8}(g_1^2 + g_2^2)(H_1 \bar{H}_1 - H_2 \bar{H}_2)^2 - \frac{g_2^2}{2} {\lvert H_1 \bar{H}_2 \rvert}^2,
\end{eqnarray}

where $m_{12}^2$ is defined to be negative and $\epsilon_{12}=-\epsilon_{21}=-1$, with soft breaking parameters $m_1^2, m_2^2, m_{12}^2$ and $g_1, g_2$ are SU(2) and U(1)
gauge couplings, respectively. Decomposing each Higgs doublet field $H_1,2$ in terms of its components \cite{r19}, we get:   

\begin{eqnarray}
H_1=\left(
\begin{array}{c}
H_1^1 \\
H_1^2
\end{array}\right)=\left(
\begin{array}{c}
(v_1+\phi_1^0 - i \chi_1^0)/ \sqrt 2        \\
-\phi_1^-
\end{array}\right)\nonumber
\end{eqnarray}

\begin{eqnarray}
H_2=\left(
\begin{array}{c}
H_2^1 \\
H_2^2
\end{array}\right)=\left(
\begin{array}{c}
\phi_2^+        \\
(v_2+\phi_2^0 - i \chi_2^0)/ \sqrt 2
\end{array}\right)
\end{eqnarray}

The Higgs potential Eq.(\ref{eq:10}) is diagonalized by the following rotations:

\begin{eqnarray}
\label{eq:12}
\left(\begin{array}{c}
H^0 \\
h^0
\end{array}\right)=\left(
\begin{array}{cc}
\cos \alpha & \sin \alpha        \\
-\sin \alpha & \cos \alpha
\end{array}\right) \left(
\begin{array}{c}
\phi_1^0         \\
\phi_2^0 
\end{array}\right)\nonumber
\end{eqnarray}

\begin{eqnarray}
\left(\begin{array}{c}
G^0 \\
A^0
\end{array}\right)=\left(
\begin{array}{cc}
\cos \beta & \sin \beta        \\
-\sin \beta & \cos \beta
\end{array}\right) \left(
\begin{array}{c}
\chi_1^0         \\
\chi_2^0 
\end{array}\right)\nonumber
\end{eqnarray}

\begin{eqnarray}
\left(\begin{array}{c}
G^+ \\
H^+
\end{array}\right)=\left(
\begin{array}{cc}
\cos \beta & \sin \beta        \\
-\sin \beta & \cos \beta
\end{array}\right) \left(
\begin{array}{c}
\phi_1^+         \\
\phi_2^+ 
\end{array}\right)
\end{eqnarray} 
$G^0$, $G^\pm$ describe the unphysical Goldstone modes. The spectrum of physical states consists of: a light neutral CP-even state ($h^0$), a heavy neutral CP-even state ($H^0$), 
neutral CP-odd state ($A^0$), and apair of charged states ($H^\pm$). 

The masses of the gauge bosons and the electromagnetic charge are determined by:

\begin{align}
M_Z^2 &= \frac{1}{4}(g_1^2 + g_2^2)(v_1^2 + v_2^2),\nonumber\\
M_W^2 &= g_2^2(v_1^2 + v_2^2),\nonumber\\
e^2&=\frac{g_1^2g_2^2}{g_1^2+g_2^2}.
\end{align}
Thus, the potential (\ref{eq:10}) contains two independent free parameters, which can conveniently be chosen as:

\begin{eqnarray}
\label{eq:14}
\tan\beta=\frac{v_2}{v_1}, \quad   M_A^2=-m_{12}^2(\tan\beta+\cot\beta)
\end{eqnarray}
where $M_A^2$ is the mass of the $A^0$ boson. The masses of the other physical states read can be expressed in terms of Eq. (\ref{eq:14}):

\begin{align}
\label{eq:15}
m_{H^0,h^0}^2&=\frac{1}{2}\biggl[M_A^2+M_Z^2 \pm \sqrt{(M_A^2+M_Z^2)^2-4M_A^2 M_Z^2 \cos^2 2\beta}\biggr]\nonumber\\
m_{H^+}^2&=M_A^2+M_W^2,
\end{align}
and the mixing angle $\alpha$ in the ($H^0, h^0$)-system is derived as the following:

\begin{eqnarray}
\label{eq:16}
\tan2\alpha=\tan2\beta \frac{M_A^2+M_Z^2}{M_A^2-M_Z^2}, -\frac{\pi}{2}< \alpha \leq 0. 
\end{eqnarray}
Hence, masses and couplings are determined by only a single parameter more than in the SM.

The dependence on $M_A$ is symmetric under ${\tan\beta}\leftrightarrow{1/\tan\beta}$, and $m_{h^0}$ is constrained by:

\begin{eqnarray}
m_{h^0} < M_Z\cos2\beta < M_Z.                      
\end{eqnarray}
However, this simple scenario is changed when radiative corrections are taken into account.

The tree-level mass matrix $m_0$ of the neutral scalar system that represents bare mass system is diagonalized by Eqs. (\ref{eq:12}). Loop contributions to the quadratic part of the potential (neglecting the $q^2$-dependence of the diagrams) modify the mass matrix as:

\begin{eqnarray}
m_0\rightarrow m_0 + \delta m = m
\end{eqnarray}                               
Re-diagonalizing the one-loop matrix $m$ yields the corrected mass eigenvalues $m_{H^0,h^0}$, replacing Eq. (\ref{eq:15}), and an effective mixing angle $\alpha_{eff}$ instead of Eq. (\ref{eq:16}).
At the one-loop level, the free parameters and the fields of the Lagrangian are replaced by renormalized parameters and fields, and a set of counterterms as following \cite{r20}:
\begin{align}
\label{eq:19}
B_\mu & \rightarrow(Z_2^B)^{1/2} B_\mu,\nonumber\\
W_\mu^a & \rightarrow(Z_2^W)^{1/2} W_\mu^a,\nonumber\\ 
H_i & \rightarrow Z_{H_i}^{1/2} H_i, \nonumber\\
\psi_j^L & \rightarrow {(Z_L^j)}^{1/2} \psi_j^L,\nonumber\\ 
\psi_{j\sigma}^R & \rightarrow {(Z_R^{j\sigma} )}^{1/2} \psi_{j\sigma}^R, \nonumber\\
g_2 & \rightarrow Z_1^W (Z_2^W )^{-3/2} g_2,\nonumber\\
g_1 & \rightarrow Z_1^B (Z_2^B )^{-3/2} g_1,\nonumber\\
v_i & \rightarrow Z_{H_i}^{1/2} (v_i - \delta v_i),\nonumber\\
m_i^2 & \rightarrow Z_{H_i}^{-1} (m_i^2 + \delta m_i^2),\nonumber\\  
m_{12}^2 & \rightarrow Z_{H_1}^{-1/2} Z_{H_2}^{-1/2} (m_{12}^2 + \delta m_{12}^2).
\end{align}

The complete definitions and the explicit expressions of the renormalization constants of the other sectors: sfermion sector, MSSM parameters and fields including those of SM as the electric charge and the masses of $W$, $Z$, and the fermions and their counterterms in addition to $\tan \beta$, all these are treated as described in \cite{r21}, to deliver all counterterms required for propagators and vertices appearing in the amplitudes. 

\subsection{Real Photon Emission}

The soft IR divergences in the $M_{virtual}$ originate from the contributions of virtual photon exchange in loops \cite{r22}. These soft (IR) divergences can be cancelled by the real photon bremsstrahlung corrections in the soft photon limit. The real photonic emission process:
\begin{eqnarray}
e^+(p_1)+e^-(p_2)\rightarrow{\tilde{\chi}}_1^0(p_3)+{\tilde{\chi}}_1^0(p_4)+h^0(p_5)+\gamma(k_\gamma),\nonumber
\end{eqnarray}
where the photon of momentum $(k_\gamma)$ radiates from the electron/positron 𝑒$e^\pm$, can have either soft or collinear nature. The collinear singularity is regularized by keeping electron (positron) mass. The general phase-space-slicing method (PSS) is adopted to separate the soft photon emission singularity from the real photon emission processes. In the PSS approach the soft and collinear regions are excluded from phase space by appropriate phase-space cuts. By introducing an arbitrary small soft cutoff, we separate the overall integration of the $2\rightarrow4$ phase space into singular and non-singular regions by the soft photon cut off, $\Delta E=\delta_s\sqrt{s}/2$, i.e. $E_\gamma\leq\Delta E$, or hard, i.e. $E_\gamma\leq\Delta E$. The real cross section in Eq.(\ref{eq:3}) can then be written as \cite{r23}:
\begin{align}
\label{eq:20}
\sigma^{real}&=\sigma^{soft}(\Delta E)+\sigma^{hard}(\Delta E)\nonumber\\
&=\sigma^{0}(\Delta_{soft}+\Delta_{hard}),
\end{align}
where $\sigma^{soft}$ is obtained by integrating over the soft region of the photon phase space, and contains all the IR soft divergences of $\sigma^{real}$. To isolate the remaining collinear divergences from $\sigma^{hard}$, we further split the integration over the hard photon phase space according to whether the photon is $(\sigma^{coll})$ or is not $(\sigma^{non-coll})$ emitted within an angle $\theta$ with respect to the radiating particles such that $(1-\cos \theta)<\delta_c$, for an arbitrary small collinear cutoff $\delta_c$:
\begin{eqnarray}
\sigma^{hard}=\sigma^{coll}(\Delta\theta)+\sigma^{non-coll}(\Delta\theta).
\end{eqnarray} 
The hard non-collinear part of the real cross section, $\sigma^{non-coll}$, is finite and can be computed numerically, using standard Monte-Carlo techniques.

Due to the selectron exchange channels, one cannot separate off all Feynman diagrams with an additional photon attached to the tree-level diagrams to define pure “weak and QED corrections”, where we have $\sigma^{weak}=\sigma^{soft}$ and $\sigma^{QED}=\sigma^{hard}$. The energy of the radiated photon in the center of mass system frame is considered as a soft term, $\Delta_{soft}$, with radiated photon energy $k_\gamma^0<\Delta E$,  and a hard term, $\Delta_{hard}$, with $k_\gamma^0>\Delta E$ , where $k_\gamma^0 = \sqrt{{\lvert\vec{k}_\gamma\rvert^2}+m_\gamma^2}$ and $m_\gamma$ is the photon mass, which is used to regulate the IR divergences existing in the soft term. 

in Eq.(\ref{eq:20}), $\Delta E$ depends largely on the weak and QED components. The main part of the QED corrections arises from the leading logarithms $L_e \equiv \log(s/m_e^2)$, resulting from photons in the beam direction. This leads to a large dependence on the experimental cuts and detector specifications. Therefore, we extract the $\Delta E$ and $L_e$ terms, caused by collinear soft photon emission, from the weak corrections and add them to the QED corrections such that both corrections are now cutoff independent \cite{r24}. Now, Eq.(\ref{eq:2}) can be written to include weak and QED corrections. The total renormalized cross section $\sigma^{total}$ is expressed as:
\begin{eqnarray}
\sigma^{total} =\sigma^0 + \sigma^{virt} + \sigma^{weak} + \sigma^{QED},
\end{eqnarray}
The integrated cross section at the one-loop level, can be written in the following way:
\begin{eqnarray}
\sigma^{total} =\sigma^0 + \sigma^0 \Delta,
\end{eqnarray}
where $\Delta$, the relative correction, is given by:
\begin{eqnarray}
\Delta = \left(\sigma^{total}-\sigma^0\right)/\sigma^0,
\end{eqnarray}
which can be decomposed into the following parts, indicating their origin:
\begin{eqnarray}
\Delta = \Delta_{self} + \Delta_{vert} + \Delta_{box} + \Delta_{QED} + \Delta_{weak}.
\end{eqnarray}

\section{Numerical Results}
\label{sec:2}

In present work, two different scenarios are studied. In the higgsino scenario, the neutralinos are higgsino-like as $\mu \ll M_1,M_2$ and the process is dominated by the s-channel $Z_0$  exchange. In the gaugino scenario the neutralinos are bino-like as $\mu \gg M_1,M_2$ the selectron exchange diagrams play the most important role. The renormalization scale is taken to be  $Q=2m_{\tilde \chi_1^0}+m_{h^0}$. The SM input parameters are set as the following: \\
$\alpha (M_Z)=1/127.922$, $M_e=0.511$ MeV, $M_W=80.399$ GeV, $M_Z=91.189$ GeV, $M_t=174.3$ GeV, $M_b=4.7$ GeV.

The mass spectrum of the SUSY particles are set using two programs; Isajet, which is Monte Carlo program that simulates $e^+ e^-$ interaction, and SuSpect, which is Fortran code that calculates the supersymmetric and Higgs particle spectrum in MSSM. The free parameters that have been used in our calculations are specified as follows:
\begin{itemize}
	\item All trilinear couplings are set to a common value $A_f \left(A_{tau}=A_b=A_t\right)$, and all soft SUSY breaking parameters are assumed equal. 
	\item The MSSM Higgs sector is parametrized by the CP-odd mass, $m_A$, and $\tan\beta$.
	\item The mixing between sfermion generations is neglected, $M_{SUSY}\equiv \tilde{M}_L \simeq \tilde{M}_R$.									
\end{itemize}

\subsection{Higgsino Scenario}
The chosen SUSY parameters are set for Higgsino scenario as following:
$\tan\beta = 10$, $M_2 = 400$ GeV, $\mu = -100$ GeV, $A_f = 400$ GeV, $m_A = 700$ GeV, $M_{SUSY} = 350$ GeV. The supersymmetric mass spectrum for Higgsino scenario using Isajet and SuSpect programs are set as shown in Table ~\ref{parset1}.

\begin{figure}[H]
\includegraphics{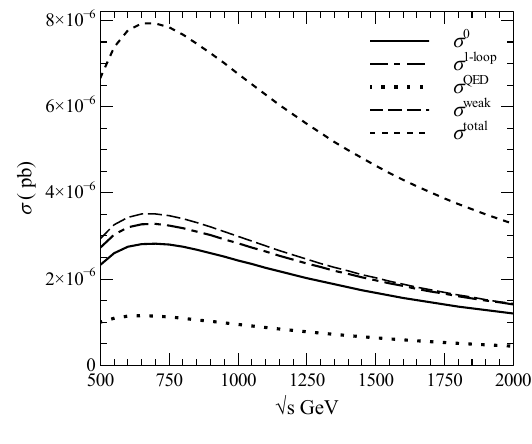}
\caption{\label{fig:5}Total cross section as a function of $\sqrt{s}$ in the
Higgsino scenario.}
\end{figure}

\begin{figure}[H]
\includegraphics{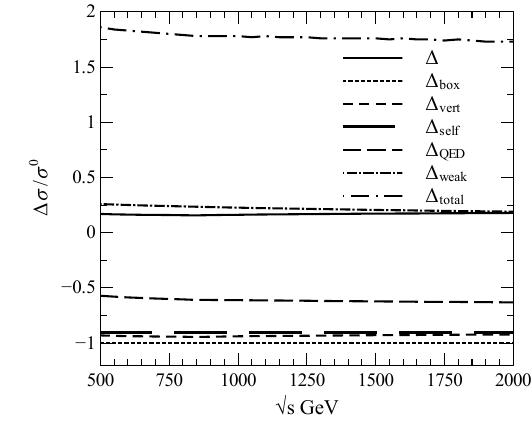}
\caption{\label{fig:6}Relative corrections as a function of $\sqrt{s}$ in
the Higgsino scenario.}
\end{figure}

Studying the dependency of the cross section on center of mass energy $\sqrt{s}$ in Fig.~\ref{fig:5} shows that the weak corrections has the maximum contribution to the total cross section, while the QED correction has the minimum contribution. Fig.~\ref{fig:6} provides detailed study of the relation between the relative corrections of the cross section and $\sqrt{s}$. The highest value of the relative correction in the virtual part is due to the self-energy contribution. The specific values of the maximum total cross section and the related center of mass energy are shown in Table ~\ref{parset2}.

\begin{table}[H]
\centering
\caption{The mass spectrum of the SUSY particles for Higgsino scenario}
\label{parset1}
\renewcommand{\arraystretch}{1.3}
\begin{tabular*}{\columnwidth}{ccccc}
\cline{2-5}
&Particle & \multicolumn{1}{c}{Mass/[GeV]} & Particle & \multicolumn{1}{c}{Mass/[GeV]} \\
\cline{2-5}
&$h^0$                 & 105.341     & $\tilde{\chi}_1^0$    & 86.3240 \\
&$H^0$                 & 700.275     & $\tilde{\chi}_2^0$    & 111.646 \\
&$A^0$                 & 700.000     & $\tilde{\chi}_3^0$    & 200.218 \\
&$H^\pm$               & 704.600     & $\tilde{\chi}_4^0$    & 416.025 \\\cline{4-5}
&$\tilde{g}$           & 1063.46     & $\tilde{\nu}_e$       & 344.128 \\\cline{2-3}
&$\tilde{\chi}_1^\pm$  & 99.2230     & $\tilde{\nu}_\mu$     & 344.128 \\
&$\tilde{\chi}_2^\pm$  & 416.032     & $\tilde{\nu}_\tau$    & 342.105 \\
\cline{2-5}
&$\tilde{e}_L$         & 352.582     & $\tilde{e}_R$         & 353.214 \\
&$\tilde{\mu}_L$       & 352.519     & $\tilde{\mu}_R$       & 353.277 \\
&$\tilde{\tau}_L$      & 349.346     & $\tilde{\tau}_R$      & 356.424 \\
\cline{2-5}
&$\tilde{u}_L$         & 345.881     & $\tilde{u}_R$         & 348.268 \\
&$\tilde{d}_L$         & 350.860     & $\tilde{d}_R$         & 354.925 \\
&$\tilde{c}_L$         & 345.667     & $\tilde{c}_R$         & 348.485 \\
&$\tilde{s}_L$         & 350.853     & $\tilde{s}_R$         & 354.932 \\
&$\tilde{t}_L$         & 281.995     & $\tilde{t}_R$         & 469.650 \\
&$\tilde{b}_L$         & 343.316     & $\tilde{b}_R$         & 362.288 \\
\cline{2-5}
\end{tabular*}
\end{table}

\begin{table}[H]
\centering
\caption{The maximum cross section in Higgsino scenario.}
\label{parset2}
\renewcommand{\arraystretch}{1.3}
\begin{tabular*}{\columnwidth}{@{\extracolsep{\fill}}clc@{}}
\cline{2-3}
&$(\sigma)_{max}/$Pb & $\sqrt{s}/$GeV   \\
\hline
\quad Born                 & $2.82 \times 10^{-6}$     & 700        \\
\quad 1--loop              & $3.28 \times 10^{-6}$     & 700        \\
\quad QED                  & $1.15 \times 10^{-6}$     & 650        \\
\quad Weak                 & $3.51 \times 10^{-6}$     & 625        \\
\quad Total                & $7.93 \times 10^{-5}$     & 625        \\
\hline
\end{tabular*}
\end{table}

\subsection{Gaugino Scenario}
The chosen SUSY parameters are set for Gaugino scenario as following:
 $\tan\beta = 10.2$, $M_2 = 197.6$ GeV, $\mu = 353.1$ GeV, $A_f = -100$ GeV, $m_A = 393.6$ GeV, $M_{SUSY} = 500$ GeV.\\
The supersymmetric mass spectrum for Gaugino scenario using the same programs, Isajet and SuSpect, are set as shown in Table ~\ref{parset3}.

\begin{figure}[H]
\includegraphics{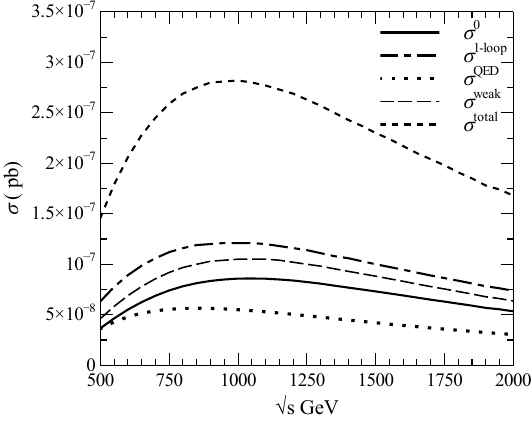}
\caption{\label{fig:7}Total cross section as a function of $\sqrt{s}$ in the
Gaugino scenario.}
\end{figure}

In this scenario, studying the dependency of the cross section on $\sqrt{s}$, Fig.~\ref{fig:7}, reveals that the one-loop correction has the highst contribution to the total cross section while the QED correction has the lowest. Table ~\ref{parset4} presents the maximum values of the total cross section and the corresponding center of mass energy.  

\begin{figure}[H]
\includegraphics{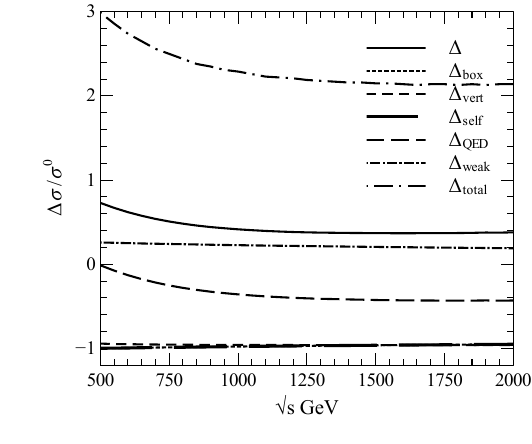}
\caption{\label{fig:8}Relative corrections as a function of $\sqrt{s}$ in
the Gaugino scenario.}
\end{figure}

In Fig. Fig.~\ref{fig:8}, the relative corrections for the three types of virtual corrections have approximately the same values in the range of $\sqrt{s} > 1000$ GeV, while $\sqrt{s} < 1000$ the virtual vertex correction has the highest value.

\begin{table}[H]
\centering
\caption{The mass spectrum of the SUSY particles for Gaugino scenario}
\label{parset3}
\renewcommand{\arraystretch}{1.3}
\begin{tabular*}{\columnwidth}{ccccc}
\cline{2-5}
&Particle & \multicolumn{1}{c}{Mass/[GeV]} & Particle & \multicolumn{1}{c}{Mass/[GeV]} \\
\cline{2-5}
&$h^0$                 & 110.985     & $\tilde{\chi}_1^0$    & 91.5340    \\
&$H^0$                 & 393.987     & $\tilde{\chi}_2^0$    & 181.009    \\
&$A^0$                 & 393.600     & $\tilde{\chi}_3^0$    & 359.502    \\
&$H^\pm$               & 401.727     & $\tilde{\chi}_4^0$    & 378.874    \\\cline{4-5}
&$\tilde{g}$           & 525.351     & $\tilde{\nu}_e$       & 495.905    \\\cline{2-3}
&$\tilde{\chi}_1^\pm$  & 180.516     & $\tilde{\nu}_\mu$     & 495.905    \\
&$\tilde{\chi}_2^\pm$  & 379.562     & $\tilde{\nu}_\tau$    & 493.614    \\
\cline{2-5}
&$\tilde{e}_L$         & 501.812     & $\tilde{e}_R$         & 502.257    \\
&$\tilde{\mu}_L$       & 501.586     & $\tilde{\mu}_R$       & 502.483    \\
&$\tilde{\tau}_L$      & 495.440     & $\tilde{\tau}_R$      & 508.551    \\
\cline{2-5}
&$\tilde{u}_L$         & 497.123     & $\tilde{u}_R$         & 498.787    \\
&$\tilde{d}_L$         & 500.591     & $\tilde{d}_R$         & 503.475    \\
&$\tilde{c}_L$         & 497.107     & $\tilde{c}_R$         & 498.807    \\
&$\tilde{s}_L$         & 500.557     & $\tilde{s}_R$         & 503.508    \\
&$\tilde{t}_L$         & 504.356     & $\tilde{t}_R$         & 548.374    \\
&$\tilde{b}_L$         & 484.474     & $\tilde{b}_R$         & 519.044    \\
\cline{2-5}
\end{tabular*}
\end{table}

\begin{table}[H]
\centering
\caption{The maximum cross section in Gaugino scenario.}
\label{parset4}
\renewcommand{\arraystretch}{1.3}
\begin{tabular*}{\columnwidth}{@{\extracolsep{\fill}}clc@{}}
\cline{2-3}
&$(\sigma)_{max}/$Pb & $\sqrt{s}/$GeV   \\
\hline
\quad Born                 & $8.59 \times 10^{-8}$     & 1050       \\
\quad 1--loop              & $1.21 \times 10^{-7}$     & 1000       \\
\quad QED                  & $5.65 \times 10^{-8}$     & 850        \\
\quad Weak                 & $1.05 \times 10^{-7}$     & 1050       \\
\quad Total                & $2.82 \times 10^{-7}$     & 1000       \\
\hline
\end{tabular*}
\end{table}

\section{Conclusions}
\label{sec:3}
In this paper we calculated the full electroweak radiative corrections of the total cross section to the lightest neutralino pair production with light neutral Higgs boson at electron-positron LC in the frame of MSSM. The calculations were performed in an analytical method using the FeynArts-3.6 and FormCalc-7.1 computer packages, where we modified the MSSM model file implemented in FeynArts-3.6 by adding the renomalization constants and counterterms of all MSSM particles. We have calculated the weak and QED corrections, which contribute significantly to the total cross section. 

The full electroweak radiative corrections are in the range of 173-186\% for Higgsino scenario, and of 215-300\% for Gaugino scenario, thus they have to be taken into account in future linear collider experiments. The maximum cross sections are presented in Tables ~\ref{parset2} and ~\ref{parset4} for both scenarios. In general, by comparing the cross section values of the two scenarios, it is found that Higgsino scenario has larger values for all types of correction than that of the Gaugino scenario. The complete one-loop corrections for the Gaugino scenario are in the range of 150-200\% for the same reaction according to ref. \cite{r25}, showing the effect of the chosen parameters on the calculations.

\section*{Acknowledgments}
We would like to express our sincere gratitude to our
advisor Pro. Dr. M. Khaled Hegab, and our thanks to Pro. Dr.
Samiha Abou Stiet for revising the paper. We have to express 
our appreciation to Dr. Ibrahim A. Abdul-Magead for sharing 
his fruitful thoughts.

\end{multicols}

\end{document}